\documentclass[aps,prb,twocolumn,10pt,superscriptaddress,letterpaper,showpacs,reprint,preprintnumbers]{revtex4-1}
\usepackage{color}
\usepackage{hyperref}
\usepackage{graphicx}
\usepackage{amsmath}
\usepackage{amssymb}
\usepackage{amsfonts}
\usepackage{wasysym}
\usepackage{bbm}
\usepackage{multirow}
\usepackage{float}
\usepackage{epsfig}
\usepackage{epstopdf}
\usepackage{appendix}
\allowdisplaybreaks
\renewcommand{\narrowtext}{\begin{multicols}{2} \global\columnwidth24.5pc}
\renewcommand{\widetext}{\end{multicols} \global\columnwidth42.5pc}
\newcommand{\adag}{a^{\dag}}

\newcommand{\sx}{\sigma_{x}}

\newcommand{\sy}{\sigma_{y}}

\newcommand{\sz}{\sigma_{z}}

\newcommand{\taux}{\tau_{x}}
\newcommand{\tauy}{\tau_{y}}
\newcommand{\tauz}{\tau_{z}}

\newcommand{\ky}{k_{y}}
\newcommand{\ko}{k_{0}}
\newcommand{\at}{\tilde{a}}
\newcommand{\atdag}{\tilde{a}^{\dag}}
\newcommand{\ate}{\tilde{a}_{NL}}

\newcommand{\bchi}{\bar{\chi}}

\newcommand{\DD}{|\Delta|}
\newcommand{\TP}{t_{\perp}}
\newcommand{\DP}{\Delta_{\perp}}

\newcommand{\PL}{P_{L}}
\newcommand{\PSL}{P_{\Omega_{S},L}}
\newcommand{\TL}{T_{L}}

\newcommand{\PSR}{P_{\Omega_{S},R}}

\newcommand{\PS}{P_{\Omega_{S}}}
\newcommand{\TS}{T_{\Omega_{S}}}
\newcommand{\TE}{T_{\Omega_{E}}}
\newcommand{\PE}{P_{\Omega_{E}}}

\newcommand{\bEta}{\bar{\eta}}
\newcommand{\bDelta}{\bar{\delta}}

\newcommand{\oL}{O_{L}} 
\newcommand{\oR}{O_{R}} 
\newcommand{\lang}{\langle}
\newcommand{\rang}{\rangle}
\newcommand{\OmS}{\Omega_{S}}
\newcommand{\OmE}{\Omega_{E}}
\newcommand{\STOP}{S_{top}}
\newcommand{\STRI}{S_{triv}}
\newcommand{\ZN}{\mathbb{Z}_{N}}
\newcommand{\Ztwo}{\mathbb{Z}_{2}}
\newcommand{\Zone}{\mathbb{Z}_{1}}
\newcommand{\ZZ}{\mathbb{Z}}

\makeatletter
\def\p@subsection{}
\makeatother
\begin{document}

\title{Detecting Majorana fermions in quasi-one-dimensional topological phases using nonlocal order parameters}
\author{Yasaman Bahri}
\affiliation{Department of Physics, University of California at Berkeley,
Berkeley, CA 94720, USA}
\author{Ashvin Vishwanath}
\affiliation{Department of Physics, University of California at Berkeley,
Berkeley, CA 94720, USA}

\begin{abstract}
Topological phases which host Majorana fermions can not be identified via local order parameters. We give simple nonlocal order parameters to distinguish quasi-one-dimensional (1D) topological superconductors of spinless fermions, for any interacting model in the absence of time reversal symmetry. These string or ``brane" order parameters are natural for measurements in cold atom systems using quantum gas microscopy. We propose them as a way to identify symmetry-protected topological phases of Majorana fermions in cold atom experiments via bulk rather than edge degrees of freedom. Subsequently, we study two-dimensional (2D) topological superconductors via the quasi-1D limit of coupling $N$ identical chains on the cylinder. We classify the symmetric, interacting topological phases protected by the additional $\mathbb{Z}_N$ translation symmetry. The phases include quasi-1D analogs of (i) the $p+ip$ chiral topological superconductor, which can be distinguished up to the 2D Chern number mod 2, and (ii) the 2D weak topological superconductor. We devise general rules for constructing nonlocal order parameters which distinguish the phases. These rules encode the signature of the fermionic topological phase in the symmetry properties of the terminating operators of the nonlocal string or brane. The nonlocal order parameters for some of these phases simply involve a product of the string order parameters for the individual chains. Finally, we give a physical picture of one of the topological phases as a condensate of certain defects, which motivates the form of the nonlocal order parameter and is reminiscent of higher dimensional constructions of topological phases.
\end{abstract}

\date{\today}
\maketitle

\indent Quantum phases with emergent Majorana fermion excitations have received much attention in the past several years.\cite{AliceaRev,MourikExp, DengExp, DasExp, RokhinsonExp, ChurchillExp, FinckExp} Majorana fermions are known to appear at the boundaries and topological defects of exotic one-dimensional (1D) \cite{Kitaev} and two-dimensional (2D) topological superconductors. \cite{ReadGreen,QiDIII} Cold atom realizations of such phases would serve as a new platform for studying and manipulating Majorana fermions.

\indent At the same time, a general framework for classifying quantum phases continues to be developed. Important achievements include the classification of free fermion systems, \cite{Schnyder,KitaevKTheory} an understanding of interaction effects in certain symmetry classes, \cite{Fidkowski1,Fidkowski2,TurnerClass,BurnellClass,LukaszClass,ChenClass,WangClass,PotterClass} and general methods for many of the symmetry-protected bosonic or fermionic systems with interactions. \cite{ChenCOM,ChenCOHO,ChenScience,Schuch,GuSUPER,LuClass,VishClass,PotterClass2} In 1D, where matrix product states provide a framework for describing ground states, the gapped bosonic symmetry-protected topological phases have been completely classified. \cite{ChenCOM,Schuch} The classification extends to 1D fermions because of their equivalence with bosons.\cite{ChenCOM} A resulting question is how to distinguish such phases via accessible observables. Fully symmetric phases have no broken symmetries and hence are immune to a local order parameter description, but they nonetheless have different topological ``fingerprints."

\indent For bosons, the insight obtained from the classification of quantum phases enables design of nonlocal order parameters that extract the defining quantities, associated with the cohomology group of the symmetry group, which characterize a bosonic symmetry-protected phase. This problem has been fully addressed in 1D in Refs. \citenum{Haegeman}, \citenum{Pollmann}, while recent progress\cite{Zaletel} in two dimensions has also been made. The 1D bosonic nonlocal order parameters (OPs) of Refs. \citenum{Haegeman}, \citenum{Pollmann} are robust in that they are strictly symmetry rather than wavefunction dependent. That is, they yield a fixed value throughout an entire quantum phase. 

\indent We contrast such nonlocal OPs with conventional string order for bosonic systems, for instance of the AKLT type,\cite{denNijs,Kennedy,OshikawaInt} which arises for the 1D Haldane phase protected by $\Ztwo \times \Ztwo$ spin rotation symmetry. The latter string order measures extraneous aspects of the wavefunction besides the topological information and so yields a continuously varying value within the phase, as would local order parameters for broken symmetries. In spite of this, AKLT string order has proven useful in many contexts. Especially, nonlocal OPs of this type, those which can be very simply expressed in terms of physical site operators, are more natural candidates for measurements in experiments.\cite{Endres}

\indent General nonlocal OPs for fermionic topological phases, that is, order parameters which go beyond a specific model to distinguish an entire quantum phase, have not, to our knowledge, been studied in any dimension. We address this problem for quasi-1D fermions by constructing string or ``brane"\cite{PerezGarciaSO} nonlocal OPs analogous to AKLT string order. Our basic building block will be the Majorana chain. Attempts at extending string order to brane order (an order parameter covering an area rather than a line) for systems beyond a single chain have been discussed before,\cite{AnfusoFragility,PerezGarciaSO} in particular in the context of the Haldane phase.\cite{BergRiseFall} 

\indent The structure of the paper is as follows. We first consider (\textbf{Section \ref{sec:SecI}}) quasi-1D spinless fermion topological superconductors with interactions but no symmetries. It is known that AKLT string order can be derived via a nonlocal mapping which transforms the topological Haldane phase into a system with broken spin rotation symmetries. \cite{Kennedy,OshikawaInt,PollmannSP} Likewise, the nonlocal Jordan-Wigner mapping can transform certain fermionic topological properties to the broken symmetry order of a bosonic system. This is one way to obtain nonlocal OPs distinguishing the fermionic topological and trivial phases of our system. The bulk of these order parameters measures the fermion parity of each site ($e^{i\pi n_{i}}$, where $n_{i}$ is the fermion occupation of site $i$), while their terminating operators may either be fermionic or bosonic.

\indent Our order parameters are relevant for cold atom experiments (\textbf{Section \ref{sec:SecII}}), which have seen recent breakthroughs with the development of the quantum gas microscope, \cite{BakrQGM,Bakr} as well as subsequent measurements of nonlocal order,\cite{Endres} by the groups of Greiner and Bloch, respectively. By making simultaneous measurements on all lattice sites of, for instance, the particle parity, these experiments constitute a nonlocal probe of the many-body system that is particularly well suited to identifying topological phases. In contrast, most other probes measure local properties, such as correlations between a pair of local operators, which makes them blind to the subtle order in the bulk of topological phases. Measuring topological aspects of free fermion band structures in cold atom systems has been discussed; \cite{AlbaExp,ZhaoExp,GoldmanExp,PriceExp,AtalaExp} here, however, we will be concerned with generic interacting systems, in particular topological superconductors. As an example of a quantity accessible with current experimental techniques, we describe a system of two identical chains for which a topological phase can be detected via measurements of fermion parity alone. 
 
\indent In \textbf{Section \ref{sec:SecIII}}, we add a protecting $\ZN$ translation symmetry to $N$ identical chains as studied in Section \ref{sec:SecI}. We describe the symmetric, interacting phases (listed in \textbf{Table \ref{tab:Correspondence}}), which capture some interesting 2D phases in a quasi-1D setting. The classification distinguishes certain topological indices in the case of free fermions. The topological ``fingerprint" of the quantum phases (\textbf{Section \ref{sec:SecIV}}) can be encoded in simple symmetry transformation rules obeyed by local operators terminating the edges of the nonlocal string or brane in the order parameter (\textbf{Table \ref{tab:EvenOddTable}}). Following Ref. \citenum{Pollmann} for bosonic systems, we refer to these as ``selection rules." For the symmetry class of our interest, these rules uniquely distinguish the symmetric phases. We conclude by describing one of the fermionic topological phases and its order parameter selection rule in terms of a bosonic model with condensed composite objects formed from Ising defects.   

\indent Because of the 1D correspondence of bosonic and fermionic systems, our results can be supported by working in either set of variables. In the main text, we mainly take the bosonic point of view and discuss the fermionic description in \textbf{Appendix \ref{sec:AppII}}. In \textbf{Appendix \ref{sec:AppIII}}, we outline a derivation of selection rules for fermionic nonlocal OPs. Understanding the rules in fermionic variables directly may be relevant for constructing order parameters for higher dimensional fermionic systems, beyond the regime in which bosons and fermions are equivalent. Throughout the paper, we use the term nonlocal OP to include string and brane order.  
\section{Interacting Spinless Fermion Topological Superconductors}
\label{sec:SecI}

\subsection{Example: Single Majorana chain}
\indent To illustrate the general form of the order parameters, we first consider Kitaev's spinless p-wave topological superconductor on an open chain with Hamiltonian:
\begin{subequations}
\label{eq:HKitaev}
\begin{align}
& H_{Kit} = \sum_{i} (-t\adag_{i}a_{i+1} + \DD a_{i}a_{i+1} + h.c.) - \mu (\adag_{i}a_{i} - \frac{1}{2})\\
&= \frac{i}{2} \sum_{i} \left[ (-t+\DD)\chi_{i}\bchi_{i+1} + (t+\DD)\bchi_{i}\chi_{i+1} - \mu \chi_{i} \bchi_{i}\right] 
\end{align}
\end{subequations}

\noindent with site fermion operators $a_{i} = \frac{1}{2}(\chi_{i}+i\bchi_{i})$ and Majorana operators $\chi_{i}, \bchi_{i}$. \cite{Kitaev} The phase of the superconducting order parameter $\Delta=\DD e^{i\theta}$ has been gauged away. For $|\frac{\mu}{2t}|<1$, there are gapped topological phases if $\DD \neq 0$ and a gapless normal phase if $\DD =0$; if $|\frac{\mu}{2t}|>1$ for any $\DD$, the phase is gapped and trivial. Let the Jordan-Wigner mapping be $\sigma^{x}_{i} = e^{i\pi n_{i}}, \sigma^{y}_{i} = \prod_{j<i}e^{i\pi n_{j}} \bchi_{i}, \sigma^{z}_{i} = - \prod_{j<i} e^{i\pi n_{j}} \chi_{i}$. The Majorana chain maps onto an XY-type spin model in a transverse magnetic field. Fermion parity $\prod_{i} e^{i\pi n_{i}}$, which implements $a_{i} \rightarrow -a_{i}$, corresponds to a $\Ztwo$ spin symmetry $\prod_{i} \sigma^{x}_{i}$ via the Jordan-Wigner mapping. This symmetry is broken or unbroken, respectively, in the spin model when the corresponding fermionic model is in a topological or trivial phase. This is in fact a general correspondence between 1D fermionic and bosonic systems, which we will discuss shortly.

\indent Consider an Ising limit (e.g. set $t=\DD$) of the spin Hamiltonian obtained from the Kitaev model: $H_{spin} = \sum_{i} -J(\sigma^{z}_{i}\sigma^{z}_{i+1} + g \sigma^{x}_{i})$ with $J=\DD$ and $g=-\frac{\mu}{2\DD}$. A two-point spin correlation function is nonzero in the spin ordered phase and vanishes in the disordered phase. It maps to a string OP which distinguishes the topological from the trivial phase in this limit: 
\begin{align}
\langle \sigma^{z}_{i} \sigma^{z}_{k} \rangle = \langle (-i\bchi_{i}) \prod_{j=i+1}^{k-1} e^{i\pi n_{j}} \chi_{k} \rangle \label{eq:Ising}
\end{align}

\noindent Note that two-point correlations are insensitive to the linear combination of states used in the ground state subspace. This is important because the fermion ground states in the topological phase are symmetric and anti-symmetric combinations of the $\Ztwo$ breaking spin ground states. This is due to a superselection rule for fermionic systems which requires fermionic states to have definite parity. 

\indent The Kitaev model Eq. \ref{eq:HKitaev} has additional symmetries, for instance time reversal $\chi \rightarrow \chi, \bchi \rightarrow -\bchi$; this constrains the possible two-point spin correlations which can be chosen. $\sigma^{z}$ or $\sigma^{y}$ correlations are nonzero in the broken symmetry regimes of $t>0$ or $t<0$, respectively, so that we have either $(\bchi_{i}, \chi_{k})$ string termination operators, as in Eq. \ref{eq:Ising}, or $(\chi_{i}, \bchi_{k})$. For fermion models with strictly no other symmetries, these constraints will not occur.

\indent We can also construct a string OP which is nonzero in the fermionic trivial phase. The self-duality of the quantum transverse Ising model under the mapping $\tau^{x}_{i+\frac{1}{2}} = \sigma^{z}_{i} \sigma^{z}_{i+1}, \tau^{z}_{i+\frac{1}{2}} = \prod_{j>i} \sigma^{x}_{j}$ to domain wall variables on bonds yields $H_{dual} = \sum_{i} (-\DD \tau^{x}_{i+\frac{1}{2}} + \frac{1}{2} \mu \tau^{z}_{i-\frac{1}{2}}\tau^{z}_{i+\frac{1}{2}})$ in the thermodynamic limit ($t=\DD$). A two-point correlation in the $\tau$ variables distinguishes the two phases. \cite{AnfusoFragility} This yields a fermion string OP which is nonzero in the trivial phase and vanishes in the topological:  
\begin{align}
\langle \tau^{z}_{i+\frac{1}{2}} \tau^{z}_{k+\frac{1}{2}} \rangle \sim \langle \prod_{j} \sigma^{x}_{j} \rangle = \langle \prod_{j} e^{i\pi n_{j}} \rangle \label{eq:IsingDual}
\end{align}

\subsection{General form}
\indent We used the Kitaev model, and in particular, an Ising limit of its spin model, to illustrate a more general correspondence which holds for quasi-1D topological superconductors of spinless fermions with interactions and no symmetries. These fermionic phases were classified in Ref. \citenum{ChenCOM} by considering the bosonic phases protected by a global bosonic $\Ztwo$ symmetry corresponding to fermion parity. There are only two gapped phases possible, which are identified as the $\Ztwo$ symmetry broken or unbroken phases; via the Jordan-Wigner mapping, they correspond to fermionic symmetric phases that are, respectively, topological (with boundary Majorana zero modes) or trivial (no gapless edge modes). The models we cite have translation symmetry along their infinite dimension, which can multiply the number of possible phases by a factor,\cite{ChenCOM} but we neglect this, focusing only on topological distinctions. In this case, there are two distinct phases.

\indent We can distinguish the phases in the bosonic variables and map the result to fermions. Any two-point correlation function $\langle O_{i} O'_{j} \rangle$, with $O/O'$ local operators which are odd under the $\Ztwo$ symmetry operation, is generically nonzero as $|i-j|\rightarrow \infty$ in the the spin ordered phase and vanishes in the disordered phase. Hence, the two-point function maps to a fermionic string OP (for one chain) or brane OP (for two or more chains) whose bulk measures fermion parity and which is terminated by \emph{fermionic} operators. 

\begin{figure}[t] 
\centering
\includegraphics[width=3.1in]{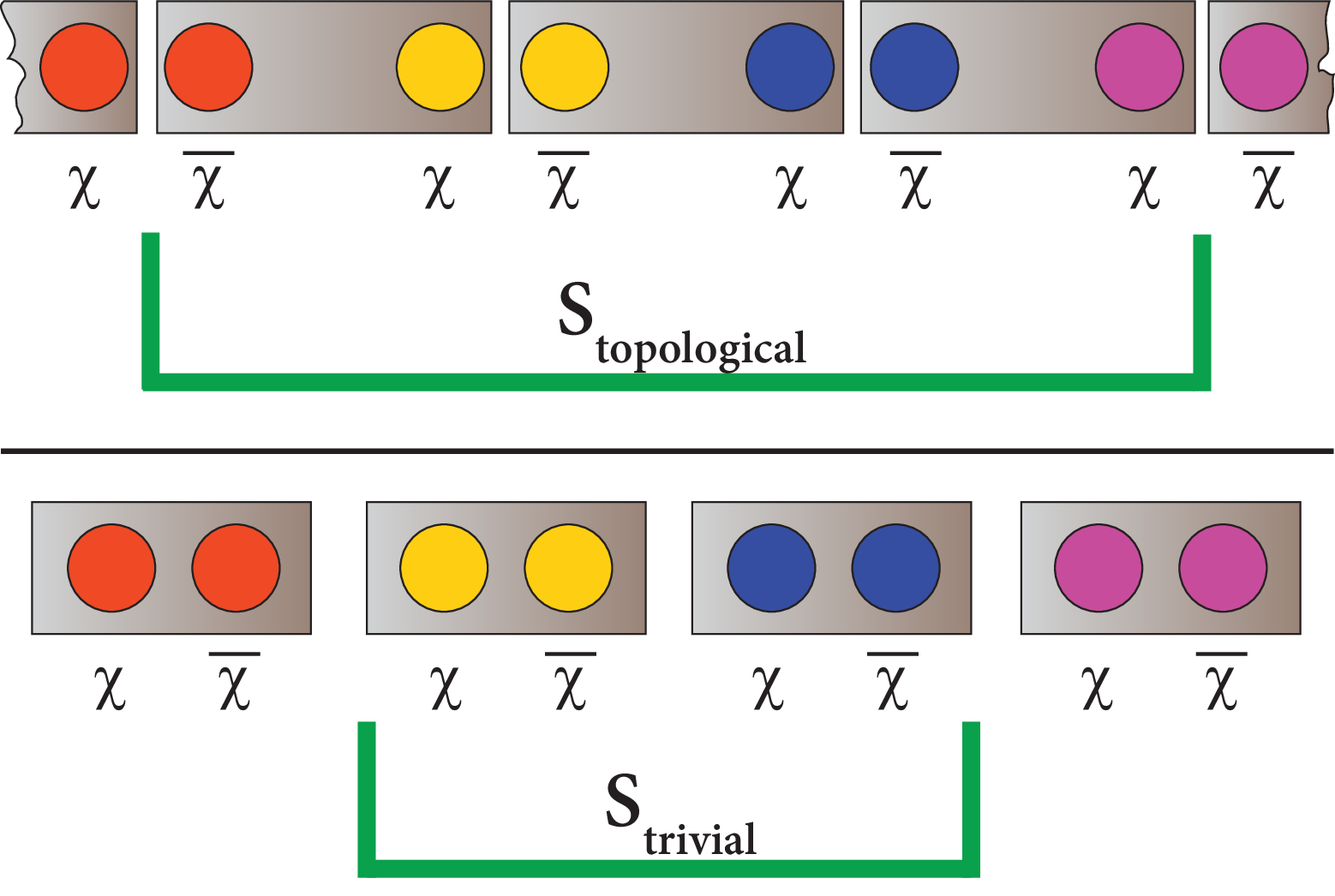} 
\caption{Single chain at top shows two Majorana fermions $\chi, \bchi$ per site (circles with same color) with nontrivial pairing (boxes). The nonlocal order parameter $\STOP$ for the topological phase measures this pairing by measuring all the operators within the bounds of the green line. It measures a ``fractional" part of the physical sites (red, purple) on the edges of the region over which it acts. Bottom chain shows a phase with trivial Majorana pairings, which is measured by $\STRI$.}
\end{figure}

\indent The spin disordered (i.e. symmetric) phase is not susceptible to local order parameters. Rather, utilizing bosonic selection rules proposed in [\!\!\citenum{Pollmann}], we conclude that a nonlocal OP which is nonzero in this phase should apply the local $\Ztwo$ symmetry over a domain in the bulk, and the domain should be terminated by operators which are $\Ztwo$ invariant. In Appendix \ref{sec:AppI}, we discuss why this order parameter vanishes in the ordered phase of spins. Mapped to fermions, an order parameter which is nonzero in the trivial phase and vanishes in the topological phase would consist of a bulk which measures fermion parity and which is terminated by local \emph{bosonic} operators. 

\indent The appearance of fermionic or bosonic terminations for a nonlocal OP is a fermionic selection rule, analogous to those described in [\!\!\citenum{Pollmann}] for bosonic systems, which distinguishes the two phases. As an alternative to using the Jordan-Wigner mapping, in Appendix \ref{sec:AppIII} we justify fermionic selection rules for nonlocal OPs from fermions directly based on ideas from fermion classification.\cite{TurnerClass} 

\indent To summarize, order parameters for the two topologically distinct phases of interacting spinless fermion topological superconductors can be constructed with the form:
\begin{subequations}
\label{eq:GeneralForm}
\begin{align}
S_{top} &= \langle O_{FL} \prod_{j \in \Omega} e^{i\pi n_{j}} O_{FR} \rangle \label{eq:GeneralFormA}\\
S_{triv} &= \langle O_{BL} \prod_{j \in \Omega} e^{i\pi n_{j}} O_{BR} \rangle \label{eq:GeneralFormB}
\end{align}
\end{subequations}
\noindent where $O_{FL/R}, O_{BL/R}$ are local fermionic or bosonic operators near the left, right edges of region $\Omega$. $\STOP$ is nonzero in the topological phase and vanishes elsewhere; the behavior of $\STRI$ is reversed. These are the \emph{generic} values, as we now discuss. 
\subsection{Remark}
\indent While the order parameters proposed throughout this paper can be used for general interacting models, their values depend in part on the state, as it requires evaluating matrix elements of certain local operators. This is no different than tailoring an order parameter for a symmetry breaking theory: certain operators may be more ``optimal" for detecting the broken symmetry because they yield larger magnitudes, while specific models may have larger symmetry groups which we can identify from the outset. For instance, the symmetry group of the quantum Ising model with $\sigma^{z}$ nearest-neighbor couplings includes time reversal of spins followed by $\pi$ rotation about y, so that $\langle \sigma^{y}_{i} \rangle = 0$.

\subsection{Microscopic picture}
\indent We explain why fermionic or bosonic terminating operators distinguish the topologically distinct fermionic phases. To illustrate, we specialize to string order in the Ising limit of the single Kitaev chain. Introduce bond fermions $\at_{i} = \frac{1}{2}(\chi_{i+1} + i\bchi_{i})$ of re-paired Majoranas, neglecting the nonlocal fermion $\ate = \frac{1}{2}(\chi_{1} + i\bchi_{N})$ by working on an infinite chain. This basis exactly solves the $t=\DD, \mu=0$ limit. The topological and trivial phase string OPs \eqref{eq:Ising},\eqref{eq:IsingDual} can be rewritten ($k \geq i+1$):
\begin{subequations}
\begin{align}
\STOP &= (-i\bchi_{i}) \prod_{j=i+1}^{k-1} e^{i\pi n_{j}} \chi_{k} \propto \prod_{j=i}^{k-1} e^{i\pi \tilde{n}_{j}} \\
\STRI &= \prod_{j=i}^{k} e^{i\pi n_{j}} \propto (\at_{i-1} + \atdag_{i-1}) \prod_{j=i}^{k-1} e^{i\pi \tilde{n}_{j}} (\at_{k} - \atdag_{k}) 
\end{align}
\end{subequations}

\indent Evidently, the fermionic or bosonic nature of the terminations depends on the basis used. The topological ground states at $t=\DD, \mu=0$ have uniform bulk filling in the \emph{bond} fermion basis $\at_{i}$ so that $\STOP$, which measures their parity, is nonzero. A weak perturbation $\mu \neq 0$ drives the system away from this ``bond-centered" ordering. It favors on-site Majorana pairings and, in perturbation theory, create localized pairs of bond fermion ``defects" with respect to the unperturbed state. Because the defects come in pairs and are localized, they only weakly modify the bond fermion parity $\STOP$ measured in a region. They are more likely to fall into the bulk region of $\STOP$, in which case they do not modify the bond fermion parity measured, rather than cross its ends. Hence, the value of $\STOP$ remains nonzero in the topological phase. A dual picture holds for site fermions $a_{i}$ deep in the trivial phase, for which site fermions are a good basis to use (i.e. the wavefunction is simple in this basis). This explains the nonzero value of $\STRI$ in the trivial phase. 

\indent On the other hand, such string OPs vanish in the complementary phases. To understand how this occurs, consider perturbatively evaluating $\STOP$ in the trivial phase of the $t=\DD$ Kitaev model. The ground state at the point $H_{0} = \frac{\mu}{2} \sum_{i} e^{i\pi n_{i}}$ with $\mu < 0$ is the site fermion vacuum $|0\rangle$ which is then corrected by the perturbation $V=\DD \sum_{i} i\bchi_{i} \chi_{i+1} = \DD \sum_{i} (a_{i} - \adag_{i})(a_{i+1} + \adag_{i+1})$. $V$ corrects $|0\rangle$ by creating localized pairs of fermion ``defects" relative to $|0\rangle$; these pairs delocalize, or new ones are created, with higher orders of perturbation theory. We see that $V$ preserves fermion parity not just globally but also ``locally," in a certain sense; locality is a strong constraint on physically allowed operators. On the other hand, $\STOP$ connects states which differ in \emph{site} occupation only at two widely separated points $i, k$. Such states cannot arise through the effects of a local \emph{and} fermion parity preserving perturbation applied to an initial state with uniform occupation throughout. Hence, $\STOP$ should remain zero away from the point $H_{0}$ as $|i-k|\rightarrow \infty$, and this holds for the entire phase. $\STRI$ likewise vanishes in the topological phase using a similar argument.

\section{Nonlocal Order and Quantum Gas Microscopy in Cold Atoms}
\label{sec:SecII}

\indent Enabled by advances in single-site resolved imaging of optical lattices,\cite{BakrQGM,Bakr} nonlocal measurements in cold atom systems are now possible and were recently demonstrated\cite{Endres} for string order in bosonic Mott insulators. \cite{BergRiseFall,DallaTorreHidden} Similarly,
\begin{align}
\STRI &= \prod_{j \in \Omega} e^{i\pi n_{j}}
\label{eq:TriParity}
\end{align}

\noindent yields a nonzero value in the trivial phase and can be measured with current experimental techniques.

\begin{figure}[b] 
\centering
\includegraphics[width=3.1in]{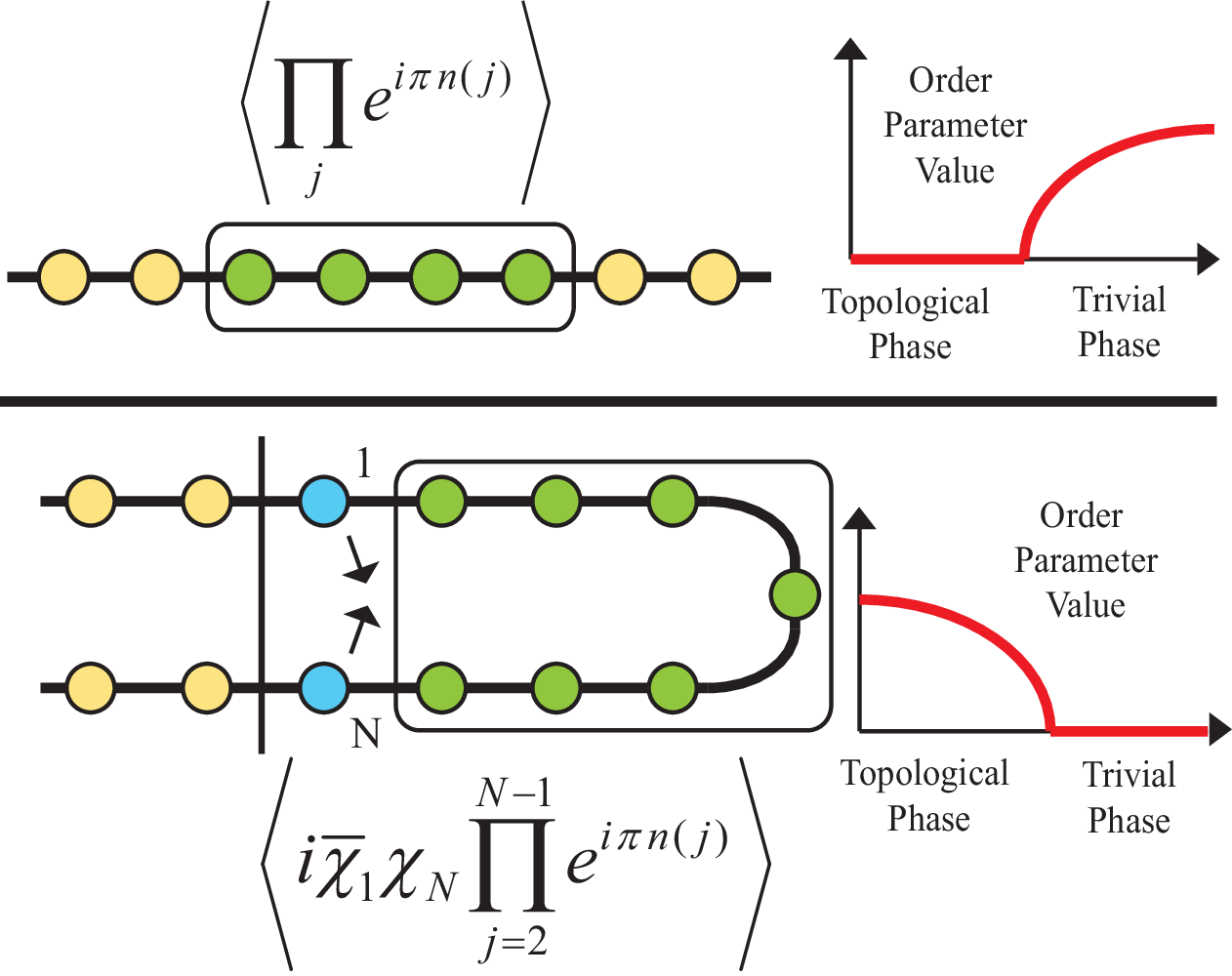} 
\caption{Top shows single Majorana chain (sites are circles) and the order parameter Eq. \ref{eq:TriParity} which is nonzero in the trivial phase and can be measured in current cold atom experiments since it only involves fermion parity (boxed green circles). Bottom shows a potential scheme for measuring order parameters with fermionic terminations such as Eq. \ref{eq:TopParity}. Fermion parity is still measured in the bulk (green), but additional measurements for the end sites labeled 1, $N$ (blue circles) must be made to extract the string order parameter value (see text).}
\label{eq:MajSiteMeasure}
\end{figure}

\indent We consider how one might measure OPs with more complex terminations. For instance, 
\begin{align}
\STOP &= (-i\bchi_{1}) \prod_{j=2}^{N-1} e^{i\pi n_{j}} \chi_{N}
\label{eq:TopParity}
\end{align}

\noindent directly detects the topological phase by generically yielding a nonzero value. The difficulty with measuring nonlocal OPs such as $\STOP$ is that they are off-diagonal in the site fermion basis imaged in experiments. We suggest a scheme for measuring a string OP such as $\STOP$ on the interval $\left[ 1, N \right]$ in the bulk of a long Majorana chain.  The idea is that by evolving the ground state in a controlled manner, such as with a tunneling Hamiltonian, we may extract the additional information needed to reconstruct the string OP (Fig. \ref{eq:MajSiteMeasure}).

\indent For instance, let a Kitaev chain ground state be $|\psi \rangle = \sum_{ijk} \beta_{ijk} |n^{I}_{i}\rangle|n^{O}_{j}\rangle |n_{k} \rangle$. Here, $|n^{I}_{i}\rangle$ is a site fermion configuration indexed by $i$ for the inner region sites $2$ to $N-1$, $|n^{O}_{j}\rangle$ indexes states for the region outside $\left[ 1, N \right]$, while $|n_{k} \rangle$ is a configuration for the string end sites $1$, $N$, with $\{ |n_{k} \rangle \}_{k=1}^{4} = \{ |0\rangle, a^{\dag}_{1}|0\rangle, a^{\dag}_{N} |0\rangle, a^{\dag}_{N} a^{\dag}_{1} |0\rangle \}$. The measured value is 
\begin{align}
\langle \STOP \rangle &= \sum_{ij} 2 P_{i} \left[ -Re(\beta_{ij1} \bar{\beta}_{ij4}) + Re(\beta_{ij2}\bar{\beta}_{ij3})\right]
\end{align}

\noindent where $P_{i}$ is the parity of configuration $i$ for sites $\left[ 2, N-1 \right]$. The additional information needed beyond amplitudes $|\beta_{ijk}|$ in order to reconstruct the expectation value are certain relative phases, such as those in $\beta_{ij1} \bar{\beta}_{ij4}$ and $\beta_{ij2}\bar{\beta}_{ij3}$. 

\indent We imagine consistently starting the system in a fixed Majorana chain ground state $|\psi\rangle$. A tunneling Hamiltonian $H_{T}$ which for instance couples only sites $1$, $N$ is turned on rapidly, preserving the state. We may consider changing the experimental geometry to have the single chain folded into two in order to couple $1$, $N$. After dynamic evolution with $H_{T}$, the site fermion occupations are measured at specified times. This information, along with accurate knowledge of the Hamiltonian parameters and amplitudes $|\beta_{ijk}|$ determined from repeated measurements, would enable extraction of the necessary relative phases and reconstruction of the string OP value. 

\indent The practicality of the suggested scheme for current systems remains to be determined. A general challenge appears to be the number of measurements needed, as a ground state for $N$ sites in the deepest regions of the topological phase consists of an exponential in $N$ number of states in the site fermion basis, all with equal magnitude weights. Design of a detailed protocol to enable extraction of the off-diagonal interference terms would be interesting and will be left to future work.

\subsection{Example: Two identical chains}
\indent We consider the case of two identical Kitaev chains A, B each with parameters $(t,\DD,\mu)$ and coupled with an interchain hopping $\TP$. This high symmetry model is a special case of the general $N$ chain system with $\ZN$ translational symmetry considered in Sec. \ref{sec:SecIII}, \ref{sec:SecIV}, but we emphasize it here because of its potential experimental relevance.

\begin{figure}[b] 
\centering
\includegraphics[width=3.2in]{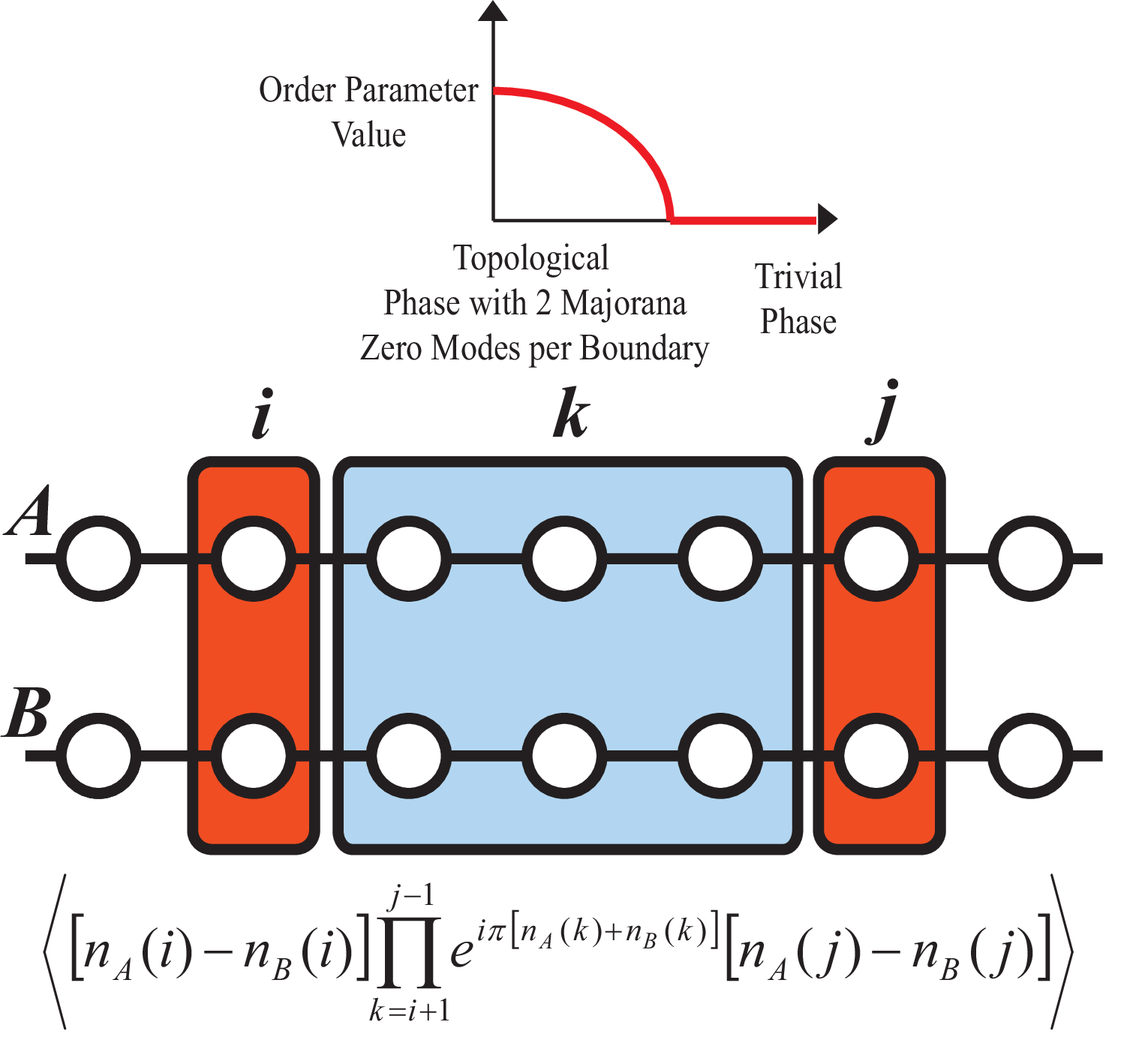} 
\caption{Geometry of the nonlocal order parameter Eq. \ref{eq:TwoChainAB} or \ref{eq:ParityDiff} for a system of two identical chains A, B (white circles are fermion sites). Bulk (blue) of the order parameter measures fermion parity while specially chosen terminating operators act on sites of the two chains (orange rectangles) separated by a large distance $|i-j|$. For instance, these terminating operators can be taken to be the fermion number difference $n_{A}-n_{B}$ (see text). This nonlocal order parameter uniquely identifies the nontrivial phase with two Majorana zero modes per boundary which is protected by the exchange symmetry.}
\end{figure}

\indent The phases of this system can be easily seen by switching to momentum $\ky=0,\pi$ in the transverse direction. The resulting Hamiltonian consists of two decoupled Kitaev models for the $\ky=0,\pi$ degrees of freedom (DOF) $\{ a_{0}(i) \}_{i} \cup \{ a_{\pi}(i) \}_{i}$ with modified chemical potentials $\mu_{\pm} = \mu \pm 2\TP$:
\begin{subequations}
\label{eq:TwoChainHami}
\begin{align}
H &= H_{Kit, A} + H_{Kit, B} - 2 \TP \sum_{i} (a^{\dag}_{iA} a_{iB} + h.c.)\\
&= H_{Kit, 0}(\mu_{+}) + H_{Kit, \pi} (\mu_{-}) 
\end{align}
\end{subequations}

\indent The phases of the system for $\DD \neq 0$ have two, one, or zero Majorana zero modes per edge as the interchain coupling $\TP$ is increased (the phase boundaries are the same as those in Fig. \ref{fig:NLOP_plot}). The phase with two Majorana zero modes per edge is protected by the exchange symmetry. To distinguish the phases we need only independently test whether the $\ky=0,\pi$ DOF are in the topological or trivial phases using Kitaev model string OPs. In regimes where only one of $\ky=0,\pi$ DOF are in the topological phase, we use string termination operators such as $\chi_{0}/\chi_{\pi} \sim \chi_{A} \pm \chi_{B}$ or those built out of $\bchi$ operators. When both $\ky=0,\pi$ DOF are in the topological phase, the terminating operators are for instance $\chi_{0}\chi_{\pi} \sim \chi_{A} \chi_{B}$. In other words, two copies of the topological phase OP of the Kitaev model, one each for the $\ky=0,\pi$ momentum DOF, detects the weakly coupled regime of this two chain system, in which each end has two Majorana zero modes. This is equivalent to a product of topological string OPs for each chain:
\begin{equation}
\begin{split}
& \langle \bchi_{0}(i)\bchi_{\pi}(i) \prod_{j=i+1}^{k-1} e^{i\pi \left[ n_{0}(j) + n_{\pi}(j) \right]} \chi_{0}(k) \chi_{\pi}(k) \rangle \rightarrow  \\ 
& \langle \bchi_{A}(i)\bchi_{B}(i) \prod_{j=i+1}^{k-1} e^{i\pi \left[ n_{A}(j) + n_{B}(j) \right]} \chi_{A}(k) \chi_{B}(k) \rangle
\label{eq:TwoChainAB}
\end{split}
\end{equation}

\noindent Taking products of string order parameter works here because of the additional protecting symmetry.

\indent We ask whether it is possible to devise an order parameter for the two chain system which involves only fermion parity but which nonetheless detects a nontrivial phase with protected Majorana zero modes. That this might be possible is suggested by the form of Eq. \ref{eq:TwoChainAB}, in which the brane is terminated by bosonic operators such as $\bchi_{A}\bchi_{B}$ rather than a fermionic operator. In fact, in Sec. \ref{sec:SecIV} we will give selection rules which the terminating operators of a nonlocal OP should obey in order to uniquely detect a symmetric phase among other symmetric phases (see Table \ref{tab:EvenOddTable}). For the phase with two Majorana zero modes per edge, the terminating operator should be bosonic (even under parity) but odd under exchange symmetry. Operators with other transformation rules under the symmetries (fermion parity and translation) detect the other symmetric phases. Therefore, 
\begin{equation}
\begin{split}
\langle \left[n_{A}(i)-n_{B}(i)\right] \prod_{j=i+1}^{k-1} e^{i\pi \left[ n_{A}(j) + n_{B}(j) \right]} 
\left[n_{A}(k)-n_{B}(k)\right] \rangle
\label{eq:ParityDiff}
\end{split}
\end{equation}
\noindent will detect the phase with two Majorana zero modes per edge. This nonlocal OP works, for instance, for a model of two identical chains with intrachain pairing and interchain diagonal hopping. While it vanishes for the special model Eq. \ref{eq:TwoChainHami} because of the model's larger symmetry group, for models with no additional symmetries this order parameter detects a topological phase.


\section{Phases with added $\ZN$ Translation Symmetry}
\label{sec:SecIII}
\subsection{$N$ chain systems on a cylinder}

\indent Consider a system of $N$ identical, interacting topological superconducting chains of spinless fermions with $\ZN$ translation symmetry transverse to the infinite chain length. The geometry is that of a cylinder with finite circumference $N$. We first seek to understand the \textbf{symmetric fermionic phases}, that is, phases with no broken symmetries. One way to identify them is to identify the corresponding bosonic phases, in part using results from the group cohomology approach to classification.\cite{ChenCOM,ChenScience} Alternatively, the fermionic phases can be identified directly by analyzing fermionic symmetry operators, as in Ref. \citenum{TurnerClass}. We do both and give a correspondence between the two descriptions. We then illustrate with models for the phases. 

\subsection{Bosonic Classification}
\indent In Refs.\citenum{ChenCOM},\citenum{ChenCOHO}, it was shown that 1D gapped bosonic phases with local interactions are in correspondence with the unbroken subgroups $G'$ of symmetry group $G$ and their second cohomology group $H^{2}(G',U(1))$. That is, given the symmetry group $G$ of a bosonic phase, its symmetries are either broken or unbroken ($G'$), and the latter subgroup can have different ``symmetry-protected topological orders." $H^{2}(G',U(1))$ is also the group of equivalence classes of projective representations $U$ of $G'$ with factor systems $\omega \in U(1)$. Qualitatively, projective representations of $G'$ reproduce the group multiplication of $G'$ up to a phase, meaning $U(g_{1})U(g_{2}) = \omega(g_{1}, g_{2}) U(g_{1}g_{2})$. However there is some redundancy in these representations: projective representations $U,U'$ with factor systems $\omega, \omega'$ are equivalent if they differ by a redefinition, that is, if $U'(g) = \beta(g) U(g)$ with $\beta \in U(1)$. This can be viewed as a kind of gauge equivalence. Certain complex phases associated with a projective representation, however, are invariant under these gauge changes and are therefore characteristic of an equivalence class and of a quantum phase. These gauge-invariant quantities, which are specified by $H^{2}(G',U(1))$, distinguish the bosonic symmetry-protected topological phases.  

\indent We take a winding path in the Jordan-Wigner mapping to transform the fermionic cylinder into a 1D infinite bosonic chain. Each bosonic unit cell accounts for one fermionic cylindrical ring and contains $N$ spin-1/2 degrees of freedom (DOF). The $\ZN$ fermion translation symmetry for the circumference, $a_{i,j} \rightarrow a_{i,(j+1)\bmod{N}}$, maps to a $\ZN$ symmetry internal to the unit cell, which we note is not translation of the spins. Crucially, any fermionic system also inherently must obey a $\Ztwo$ fermion parity symmetry (defined as $a_{i,j} \rightarrow -a_{i,j}$), and this maps to a local $\Ztwo$ symmetry of spins. Hence, the bosonic system has a $\Ztwo \times \ZN$ symmetry. To obtain fully symmetric fermionic phases, we will see that the corresponding bosonic phases may be fully or only partially symmetric.

\indent The possible fully symmetric ($G'=G=\Ztwo \times \ZN$) bosonic phases are found by identifying $H^{2}(\Ztwo \times \ZN, U(1))$. The result depends on the parity of N, since $H^{2}(\Ztwo \times \ZN, U(1)) = \Ztwo$ or $\Zone$ for $N$ even or odd. To see the physical origin of this, we digress to utilize the language of matrix product states for  describing wavefunctions.

\indent In the matrix product state language,\cite{FannesNachter,Vidal2003,Vidal2007} the coefficients $C_{i_{1}i_{2}...i_{L}}$ of a wavefunction $|\psi\rangle$ in a basis $|i_{1}i_{2}...i_{L} \rangle$ are written as scalar-valued products of matrices, with each matrix indexed by $i_{k}$. For instance, for a periodic system of L sites:\cite{Vidal2007}

\begin{equation}
\begin{split}
|\psi \rangle &= \sum_{i_{1}i_{2}...i_{L}}  C_{i_{1}i_{2}...i_{L}} |i_{1}i_{2}...i_{L} \rangle \\
&= \sum_{i_{1}i_{2}...i_{L}} tr(\Gamma_{i_{1}}\Lambda\Gamma_{i_{2}}\Lambda...\Gamma_{i_{L}}\Lambda) |i_{1}i_{2}...i_{L} \rangle
\end{split}
\end{equation}

\noindent Here, $\Gamma_{i_{k}}$ is a $D \times D$ matrix for each index $i_{k}$ referencing a physical state on site $k$, while $\Lambda$ is a nonnegative D-dimensional diagonal matrix related to the entanglement contained within the wavefunction. The matrix dimensions of $\Gamma, \Lambda$ access an ``auxiliary space." For simplicity we assume translation invariance, so the $\Gamma, \Lambda$ matrices are not explicitly $k$ dependent. An advantage of the matrix product state language is that one can easily isolate parts of the wavefunction associated with a collection of sites. 

\indent Next, consider a symmetry $\Sigma(g)$ of the wavefunction, with $g \in G$. The rule\cite{PerezGarciaSO,Pollmann} for how the matrices transform is

\begin{align}
\sum_{j'} \Sigma(g)_{jj'} \Gamma_{j'} = e^{i\theta_{g}} U^{\dag}_{g} \Gamma_{j} U_{g}
\end{align}

\noindent where $\Sigma(g)_{jj'}$ is a matrix representation of $\Sigma(g)$ and $U_{g}$ is a $D \times D$ unitary matrix multiplying $\Gamma$. This rule ensures that under a global symmetry operation the wavefunction is reproduced up to a phase, as neighboring $U,U^{\dag}$ cancel.

\indent For our $\Ztwo \times \ZN$ symmetry, we have two bosonic generators which are in correspondence with the fermionic symmetry generators, fermion parity P and translation T. The projective representation of the two bosonic generators, labeled $U_{P}, U_{T}$, each have an overall phase that can be gauge fixed, $U_{P}^{2}=U_{T}^{N}=1$. Crucially, however, the complex phase in $U_{P} U_{T} = e^{i\phi} U_{T} U_{P}$ cannot be eliminated by redefinition of the matrices $U$. Moreover, it must satisfy $e^{2i\phi} = e^{Ni\phi} = 1$ because of our gauge fixing. Hence, of the two possible solutions $\phi=0, \pi$, the latter is forbidden for $N$ odd. The gauge-invariant scalar $e^{i\phi}$ is quantized and so is preserved under smooth, gap-preserving deformations to the wavefunctions. The two possible values of $e^{i\phi}$ represent two gauge-inequivalent classes of projective representations of the symmetry group $\Ztwo \times \ZN$. One can see also, that all other complex phases which cannot be gauged away are related to this one, so specifying $e^{i\phi}$ is sufficient for labeling a projective representation. Since a product state can be represented by scalar $\Gamma_{j}$ and hence scalar $U_{P}, U_{T}$, $e^{i\phi}=1$ describes the trivial phase. In contrast, $e^{i\phi}=-1$ characterizes a topologically nontrivial phase.

\indent The above analysis identifies all symmetric bosonic phases protected by $\Ztwo \times \ZN$. We must also identify some symmetry breaking bosonic phases, as they are relevant for obtaining symmetric fermionic phases. This is because, while the analog of the fermion parity symmetry P can be broken in bosonic variables, it must be restored when mapping back to fermions. It is sufficient to label these symmetry breaking bosonic phases with their unbroken symmetry subgroup $G' \subset G$. The relevant ones are the proper cyclic subgroups generated by bosonic versions of (i) translation T and, for $N$ even only, (ii) the product of fermion parity and translation, labeled PT. We denote these by $G'=\langle T \rangle$ and $G'=\langle PT \rangle$ respectively. These two classes retain enough symmetry so that, although the analog of symmetry P is broken in bosonic variables, it and all other broken bosonic symmetries in $\Ztwo \times \ZN$ are  restored in the fermion system. That is, the resulting fermionic phases are symmetric. 

\indent For a complementary description of the phases, we also apply the approach developed in Ref. \citenum{TurnerClass} for fermions directly. On a certain subspace, fermionic symmetry operators acquire an effective form ($\hat{P},\hat{T}$) consisting of two ``fractional" pieces supported on the left and right edges of the system. As with the bosonic case, it is the commutation relations of these pieces which identify the quantum phases. We elaborate on this in App. \ref{sec:AppII} for our case and mention the result here. If the fermionic symmetry operators acquire effective forms $\hat{P} \sim P_{L} P_{R}$ and  $\hat{T} \sim T_{L} T_{R}$, where $P_{L},T_{L}$ and $P_{R},T_{R}$ are left, right fractional pieces, define $\mu, \mu'$ so that $P_{L} P_{R} = e^{i\mu} P_{R} P_{L}$ and $T_{L}T_{R} = e^{i\mu'} T_{R} T_{L}$. Then $e^{i\mu}, e^{i\mu'}$ are sufficient to characterize the symmetric fermionic phases. 

\indent The correspondence between the bosonic and fermionic classifications is given in Table \ref{tab:Correspondence}. The bosonic description consists of the unbroken symmetry subgroup $G' \subset G$ and the possible symmetry-protected topological order (``trivial" or ``nontrivial"). The fermionic description consists of the commutation relations of ``fractional" pieces of fermionic symmetry operators. In general, fractionalization of fermionic symmetry P into fermionic pieces ($\mu = \pi$) means it is broken in bosonic variables.\cite{Fidkowski2} This is not true for other fermionic symmetries (such as our $\ZN$ translation) whose behavior in bosonic variables depends in part on those of parity (App. \ref{sec:AppII}).

\begin{table}[h]
\begin{tabular}{||c||c||c||c||}
\hline
Bosonic & Fermionic & G.S.D. & Physical\\
description & description &  & example\\
& $(\mu,\mu')$ & &\\
\hline
1. Trivial & $(0,0)$ & 1 & Trivial\\
symmetric $G'=G$ & & &\\
\hline
2. Nontrivial & $(0,\pi)$ & 4 & Weak Top.\\
symmetric $G'=G$ & & & Supercond.\\
\hline
3. Symmetry breaking & $(\pi,0)$ & 2 & Strong Top.\\
$G'=\langle T \rangle$ & & & Supercond.\\
\hline
4. Symmetry breaking & $(\pi,\pi)$ & 2 & Strong Top.\\
$G'=\langle PT \rangle$ & & & Supercond.\\
\hline
\end{tabular}
\caption{Quasi-1D symmetric fermionic phases, henceforth labeled 1-4, for \textbf{N even} with symmetry group G=$\Ztwo \times \ZN$. Their descriptions in terms of both bosonic and fermionic variables are given. Ground state degeneracy (G.S.D.) listed is for a generic system (no additional symmetries) with open boundary conditions. For \textbf{N odd} only Classes 1, 3 exist.} \label{tab:Correspondence}
\end{table}

\subsection{Representative models for symmetric fermionic phases}
\indent	We identify example models for each quantum phase by considering the case when fermionic symmetry operators take simple effective forms obeying Column 2 of Table \ref{tab:Correspondence}.

\indent The fermionic operators P and T are: 
\begin{align}
P &=\prod_{i, \ky} e^{i\pi n_{\ky}(i)} & T &=\prod_{i, \ky} e^{i \ky n_{\ky}(i)}
\end{align}

\noindent where $i, \ky$ respectively index lattice sites along the cylinder length (x) and momentum around the circumference (y). $n_{\ky}(i)$ measures the occupation of the mode $a_{\ky}(i) = \frac{1}{\sqrt{N}} \sum_{j=1}^{N} e^{i\ky j} a_{i,j}$. A decomposition into Majorana operators $\chi_{\ky}(i), \bchi_{\ky}(i)$ is $a_{\ky}(i) = \frac{1}{2}\left[ \chi_{\ky}(i) + i\bchi_{\ky}(i) \right]$.

\indent We can view the set of operators $\{ a_{\ky}(i) \}_{i}$ for fixed $\ky$ as degrees of freedom (DOF) for a single Majorana wire with open boundaries. For example, to construct a model for Class 3 (symmetry breaking bosonic phase with $G'=\langle T \rangle$), we consider fixing the ground state occupations of all the $\ky \neq 0$ DOF (i.e. by putting all $\ky \neq 0$ chains into the trivial phase) so that T will act as a scalar in the ground state subspace. We treat the $\ky = 0$ DOF instead as a topological Majorana wire. Consider: 

\begin{align}
H_{3} &= \sum_{i} i\bchi_{0}(i)\chi_{0}(i+1) + \sum_{i, \ky\neq 0} i\chi_{\ky}(i) \bchi_{\ky}(i)
\end{align}

\noindent By putting the $\ky=0$ DOF in the topological phase, a two-fold degeneracy arises from the occupation or vacancy of the nonlocal complex fermion composed of a free Majorana from each of the  left, right edges. In the ground state subspace, the only distinction between the two states under a measurement of the total fermion parity is  the parity of this single nonlocal fermion. Consequently, the effective form of fermion parity P in this subspace is $\hat{P}=i\chi_{0}(1)\bchi_{0}(L)$, while $\hat{T} =1 $ is some scalar. This yields $(\mu,\mu')=(\pi,0)$. Systems in Class 3 have a two-fold ground state degeneracy on the cylinder, which is consistent with the bosonic description since $|G/G'|=2$ with $G'=\langle T \rangle$.

\indent To construct a model for Class 4 (for even N), we simply switch the treatments of the $\ky=0, \pi$ DOF.
\begin{align}
H_{4} = \sum_{i} i\bchi_{\pi}(i)\chi_{\pi}(i+1) + \sum_{i, \ky \neq \pi} i\chi_{\ky}(i) \bchi_{\ky}(i)
\end{align}

\noindent serves as a representative. Class 4 systems on the cylinder also have two-fold ground state degeneracy generically ($|G/\langle PT \rangle| = 2$).

\indent We can construct a model for Class 2 (for even N) by placing both the $\ky=0$ and $\ky=\pi$ DOF into the topological phase and fixing the site occupations of the remaining $\ky \neq 0,\pi$ DOF. A representative model is therefore
\begin{equation}
\begin{split}
H_{2} = \sum_{i} \{ i\bchi_{\pi}(i)&\chi_{\pi}(i+1) + i\bchi_{0}(i)\chi_{0}(i+1) \}  + \\
&\sum_{i, \ky\neq 0,\pi} i\chi_{\ky}(i) \bchi_{\ky}(i)
\end{split}
\end{equation}

\noindent The effective forms of the fermionic symmetry operators are $\hat{P}=\left[ i\chi_{0}(1) \chi_{\pi}(1) \right] \left[ i\bchi_{0}(L) \bchi_{\pi}(L) \right], \hat{T}=i\chi_{\pi}(1)\bchi_{\pi}(L)$, so that $(\mu,\mu')=(0,\pi)$. This particular model has four-fold ground state degeneracy, but it is physically plausible that Majorana zero modes can be gapped out in pairs on the edges without the system undergoing a topological transition. We expect that the ground state degeneracies for Classes 2-4 in Table \ref{tab:Correspondence} are the minimal values.

\subsection{Physical models}
\indent The previous models become increasingly nonlocal for large $N$; we connect them to phases of a local non-interacting model. A nearest neighbor interchain hopping $\TP$ and, for $N>2$, a nearest neighbor interchain pairing $\DP$ are allowed by translational invariance. We consider the simple lattice $p+ip$ topological superconductor (TSC) studied in [\!\!\citenum{Asahi}]. The Hamiltonian is ($N>2$, considering $N$ even):
\begin{equation}
\begin{split}
H = \sum_{i, j=1}^{j=N} (-t\adag_{i,j} a_{i+1,j} + \DD a_{i,j} a_{i+1,j} + h.c.)\\
- \mu(n_{i,j}-\frac{1}{2}) + (\DP a_{i,j} a_{i,j+1} - \TP \adag_{i,j} a_{i,j+1} + h.c.) 
\label{eq:Hami1}
\end{split}
\end{equation}

\noindent with $i,j$ indexing sites along the cylinder length and circumference, respectively. We take fixed parameters $|Im(\DP)| > 0, Re(\DP) = 0$ and $\DD > 0$. In this case, there are transitions between phases including quasi-1D versions of the 2D weak and strong TSCs as $\frac{\TP}{\mu}, \frac{t}{\mu}$ are varied. \cite{Asahi} We introduced the experimentally relevant two chain (N=2) version of this model in Sec. \ref{sec:SecII}.

\indent The phases of the system can be seen by rewriting the Hamiltonian \eqref{eq:Hami1} as
\begin{align}
H&=H_{Kit,0}(\mu_{+}) + H_{Kit,\pi}(\mu_{-}) + H'_{Kit}
\end{align}

\noindent where $H_{Kit,0}, H_{Kit,\pi}$ are the Kitaev Hamiltonians \eqref{eq:HKitaev} with $(\chi_{0},\bchi_{0})$ and $(\chi_{\pi},\bchi_{\pi})$ Majorana DOF, respectively, and, as before, $\mu_{\pm} = \mu \pm 2\TP$. To analyze the remaining piece $H'_{Kit}$ containing all $\ky \neq 0, \pi$ DOF, we transform the Majorana basis by recombining the four Majoranas for each $\ko \equiv |\ky| \neq 0,\pi$ into $\eta_{\ko}/\bDelta_{\ko}(i) \equiv \frac{1}{\sqrt{2}} (\chi_{\ko}(i) \pm \chi_{-\ko}(i))$ and $\bEta_{\ko}/\delta_{\ko}(i) \equiv \frac{1}{\sqrt{2}} (\bchi_{-\ko}(i) \pm \bchi_{\ko}(i))$. The remaining Hamiltonian can be viewed as a collection of two Kitaev chains for each $\ko \in (0,\pi)$ with Majorana DOF $(\eta,\bEta)$ and $(\delta,\bDelta)$ and modified chemical potential $\mu_{\ko} = \mu + 2\TP \cos(\ko)$. There is an ``interchain" coupling in this basis which is proportional to $Im(\DP)$ and which gaps out the Majorana zero modes of each chain:
\begin{equation}
\begin{split}
H'_{Kit} = \sum_{\ko \in (0,\pi)}& H_{Kit,(\eta_{\ko},\bEta_{\ko})}(\mu_{\ko}) + H_{Kit,(\delta_{\ko},\bDelta_{\ko})}(\mu_{\ko})\\
 - \sum_{i, \ko \in (0,\pi)}& i Im(\DP) \sin(\ko) (\eta_{\ko}\delta_{\ko} + \bDelta_{\ko}\bEta_{\ko})(i) 
\end{split}
\end{equation}

\indent Consequently, the phases of the system are determined by whether the independent $\ky=0,\pi$ DOF are in the topological phase. The $\ky=0$ DOF are topological when the associated chemical potential is sufficiently weak $|\TP + \frac{\mu}{2}|<|t|$. Here the 1D $\Ztwo$ invariant $\nu_{\ky=0} = 1$.\cite{Asahi} Likewise, the $\ky=\pi$ DOF are topological when $\nu_{\ky=\pi} = 1$ for $|\TP - \frac{\mu}{2}| < |t|$. Hence, for weak $|\frac{\TP}{\mu}|$ and $|\frac{t}{\mu}|>\frac{1}{2}$ the system has two Majorana zero modes per edge, at $\ky=0,\pi$. This phase scales to a 2D weak TSC as $N \rightarrow \infty$ in Eq. \eqref{eq:Hami1}, as in this regime the 2D $\ZZ$ invariant (Chern number) is $\nu=0$. For intermediate values $|\TP| \sim |t|$ and weak chemical potential the system has a single Majorana zero mode per edge (either at $\ky=0$ or $\pi$) and will scale to a 2D strong TSC as $N \rightarrow \infty$ since the 2D $\ZZ$ invariant $|\nu|=1$. When interchain hopping $\TP$ dominates over intrachain hopping $|\frac{\TP}{\mu}| > |\frac{t}{\mu}| + \frac{1}{2}$, the system is a weak TSC in the x-direction. Fig. \ref{fig:NLOP_plot} gives a phase diagram.

\indent Since the three sets of DOF $\ky=0$, $\ky=\pi$, and $\ky\neq 0,\pi$ decouple, we can tune each into topological or trivial phases independently while maintaining translational invariance. For instance, if the $\ky=0$ DOF form a nontrivial state, treating this as a single chain with no other symmetries, we can find a path connecting to the model $\sum_{i} i\bchi_{0}(i)\chi_{0}(i+1)$ which preserves translation symmetry since it only involves $\ky=0$ operators. Hence, the quasi-1D phases of Eq. \ref{eq:Hami1} -- namely the 2D weak TSC associated with y-direction layering and the two strong TSCs with $(\nu_{\ky=0},\nu_{\ky=\pi}) = (1,0)$ or $(0,1)$ -- would fall into Classes 2, 3, and 4, respectively, of our classification. It appears that for free fermions our classification identifies the 1D $\Ztwo$ invariants $\nu_{\ky=0}$ and $\nu_{\ky=\pi}$ and consequently $\nu \bmod 2$ since $\nu_{\ky=0} + \nu_{\ky=\pi} = \nu \bmod 2$. \cite{Asahi}  

\indent The weak TSC with $\nu_{k_{x}} \neq 0$ for $k_{x} = 0$ or $\pi$, which is associated with layering in the x-direction, appears as a trivial phase. This phase results with strong $\TP$, but in the momentum $\ky$ basis this coupling is an effective chemical potential, $\mu_{\pm} = \mu \pm 2\TP$, which favors on-site pairing of y-momentum Majoranas, driving the $\ky=0,\pi$ Majorana chains away from nontrivial pairing. It is natural that that our classification is unable to detect the topological index associated with x-translation symmetry along the cylinder length, as only y-translation has been included.

\section{Nonlocal Order Parameters for the Symmetric Phases}
\label{sec:SecIV}

\indent	We construct nonlocal order parameters to distinguish the $\Ztwo \times \ZN$ protected symmetric fermionic phases of Sec. \ref{sec:SecIII} from each other. We consider \textbf{\emph{even N}}, which encompasses the results for odd $N$. The construction is based on general distinctions made apparent by the classifications. The bosonic point of view is used below for illustration. We always work in the 1D thermodynamic limit (infinite cylinder length), and our nonlocal order parameters span a finite size $L$ along this dimension; of course, we are interested in asymptotic values as $L \rightarrow \infty$.  

\indent The bosonic description has revealed hidden structure (the breaking of certain symmetry operators) which can be used, along with recently derived selection rules for bosonic nonlocal OPs in Ref. \citenum{Pollmann}, to identify bosonic operators which distinguish the four phases. We then map back to fermions. 

\indent In the infinite bosonic chain, there is a natural unit cell which makes the $\ZN$ translation symmetry on-site. To distinguish \emph{symmetric bosonic phases}, we chose a symmetry to apply over many unit cells of the chain (a string), and we terminate the domain with operators obeying proper symmetry transformation rules. Mapped back to fermions, the nonlocal OP consists of a cylindrical brane-type region in the bulk over which a symmetry is applied, and terminating operators reside on the domain edges. We model this general form by writing the OP as the long-distance limit of $\langle O_{L} \prod_{j=1}^{L} \Sigma_{j} O_{R}\rangle$. $O_{L}/O_{R}$ are possibly different operators acting near the left, right bosonic string (fermionic brane) edges and $\Sigma_{j}$ is a symmetry operation on a bosonic unit cell (fermionic cylindrical ring). If we were to distinguish \emph{symmetry breaking bosonic phases} via two-point correlations, a similar form would be obtained when mapped back to fermions.

\indent Alternatively, we corroborate our conclusions by working directly with fermions. In Appendix \ref{sec:AppIII}, we sketch a derivation of fermionic selection rules. These rules determine how the terminating operators of the fermion order parameters should be chosen, when fermion parity is used as the bulk symmetry, to distinguish the symmetric fermionic phases, in analogy with the bosonic derivation;\cite{Pollmann} the rules are listed in Table \ref{tab:EvenOddTable}. The even or odd transformation rule for a terminating operator under P, T symmetries distinguishes among the symmetric phases of any interacting model in this symmetry class.

\begin{table}[h]
\begin{tabular}{||c||c||c||c||}
\hline
Phase (Bosonic variables) & P Trans. & T Trans. & Example\\
\hline
1. Trivial & Even & Even & $\langle S_{1} \rangle \neq 0$\\
symmetric $G'=G$ & & &\\
\hline
2. Nontrivial & Even & Odd & $\langle S_{2} \rangle \neq 0$\\
symmetric $G'=G$ & & &\\
\hline
3. Symmetry breaking & Odd & Even & $\langle S_{3} \rangle \neq 0$\\
$G'=\langle T \rangle$ & & &\\
\hline
4. Symmetry breaking & Odd & Odd & $\langle S_{4} \rangle \neq 0$\\
$G'=\langle PT \rangle$ & & &\\
\hline
\end{tabular}
\caption{Transformation rules for $O_{L}, O_{R}$ under parity and translation which uniquely distinguish among the four symmetric fermionic phases ($N$ even) when fermion parity is used as the bulk symmetry operator. Even or odd are chosen depending on the sign of $e^{-i\mu}, e^{-i\mu'}$ (App. \ref{sec:AppIII}). The examples are order parameters which are asymptotically finite in the listed phase and vanish in the other symmetric fermionic phases (primed versions $S'$ in text also work generically).} \label{tab:EvenOddTable}
\end{table}

\subsection{Construction}
\indent Consider, as an example, an order parameter which applies the symmetry P over many bosonic unit cells spanning $\left[ 1, L \right]$ (large fermionic brane). Define 
\begin{align}
S_{1} &\equiv \prod_{\ky, i=1}^{i=L} e^{i\pi n_{\ky}(i)}
\end{align}

\noindent  The expectation $\langle S_{1} \rangle$, taken with any choice of ground state, vanishes in Classes 3, 4 which have P broken in the bosonic variables (App. \ref{sec:AppI}). For Classes 1 and 2, P remains a symmetry for the bosons. Typically, applying a symmetry over an increasingly large domain of a symmetric state would yield a nonzero answer since the state should be reproduced under action of a global symmetry. 

\indent In fact, this conclusion can be false if one stays away from the system boundary in applying the symmetry over a large domain. This is because applying a symmetry over some region creates an ``artificial" boundary, in a certain sense, and different topological phases have distinct edge states which are ``created" at this artifical edge.\cite{Pollmann} The bosonic selection rules\cite{Pollmann} tell us that the operators $O_{L,R}$ which terminate the bosonic string (fermionic brane) can be chosen to transform under symmetries in such a way as to select a quantum phase. The distinction between the two phases Classes 1 and 2 in the bosonic description is the quantity $e^{i\phi} = \pm 1$ (Sec. \ref{sec:SecIII}). $O_{L,R}$ must be even under parity (because it is the symmetry used in the bulk of the order parameter) but transform as $e^{i\phi}$ under translation in order to be finite in the quantum phase labeled by $e^{i\phi}$. The OP is guaranteed to vanish in the other symmetric bosonic phase. Mapped back to the fermionic system, the brane termination operators should be bosonic but should be even or odd under translation so that the OP is nonzero in Classes 1 or 2, respectively.

\indent $S_{1}$ is, for instance, an order parameter which is nonzero in Class 1 but vanishes in Class 2 since translation invariant operators terminate its bulk. To construct a candidate with reversed behavior, we can choose operators such as $O(i) = i\chi_{0}(i)\chi_{\pi}(i)$ or $i\bchi_{0}(i)\bchi_{\pi}(i)$, which are parity invariant but translation odd, to terminate the fermionic brane. Hence, candidate order parameters which give a nonzero value for Class 2 only include
\begin{align}
S_{2} &\equiv \bchi_{0}(1) \bchi_{\pi}(1) \prod_{\ky, i=2}^{L-1} e^{i\pi n_{\ky}(i)} \chi_{0}(L)\chi_{\pi}(L) \label{eq:S2}
\end{align}

\noindent and a similarly constructed $S'_{2}$ with fermionic ends $\chi_{0}(1) \chi_{\pi}(1)$ and $\bchi_{0}(L)\bchi_{\pi}(L)$. 

\indent Finally, we utilize the fact that certain symmetries are broken in the bosonic variables to construct OPs that are nonzero in a symmetry breaking bosonic phase but vanish elsewhere. We use two-point functions in the bosonic variables $\langle U_{i} V_{j} \rangle$ with $|i-j|\rightarrow \infty$; if $U, V$ are odd under P but even under T, the result is nonzero in Class 3 but vanishes in the other symmetric fermionic phases. Likewise, operators odd under P but even under PT yield OPs which can detect Class 4. Mapped back to fermions, a few such candidates are:

\begin{align}
S_{3} &\equiv -i\bchi_{0}(1) \prod_{\ky, i=2}^{L-1} e^{i\pi n_{\ky}(i)} \chi_{0}(L) \label{eq:S3} \\
S_{4} &\equiv -i\bchi_{\pi}(1) \prod_{\ky, i=2}^{L-1} e^{i\pi n_{\ky}(i)} \chi_{\pi}(L) \label{eq:S4}
\end{align}

\noindent or similar constructions defined as $S'_{3}, S'_{4}$, with fermionic ends $-i\chi_{0}(1), \bchi_{0}(L)$ and $-i\chi_{\pi}(1),  \bchi_{\pi}(L)$, respectively.

\subsection{Application}
\begin{figure}[t] 
\centering
\includegraphics[width=3.1in]{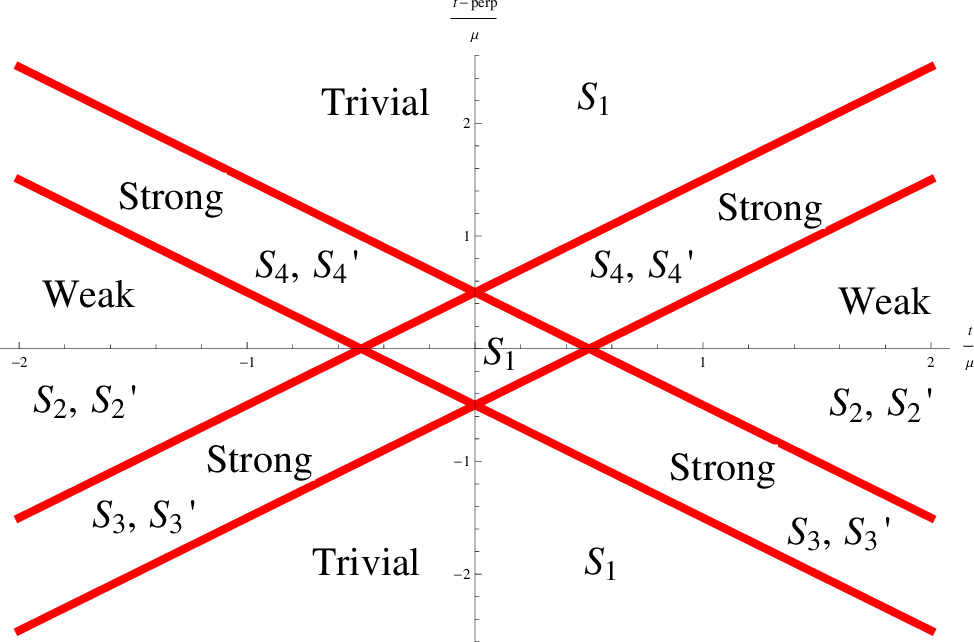} 
\caption{Phase diagram and nonlocal order parameter values for the quasi-1D $p+ip$ paired model (Eq. \ref{eq:Hami1}) with axes $(\frac{t}{\mu},\frac{\TP}{\mu})$. Phases are labeled as trivial, weak, or strong topological superconductor (center diamond is also trivial). $S/S'$ listed refer to order parameters given in text, which take on nonzero values in the phases they are associated with and vanish otherwise; unprimed, primed versions require either $t > 0$ or $< 0$, respectively.}
\label{fig:NLOP_plot}
\end{figure}

\indent We apply these order parameters to the cylinder $p + ip$ model. For instance, let us work in the $t=\DD$ limit of the strong TSC with only the $\ky =0$ DOF in the topological phase. Evaluations decouple into $\ky=0,\pi$, and $\ko \in (0,\pi)$ contributions: $\langle S_{3} \rangle = \langle \STOP \rangle_{\ky=0} \langle \STRI \rangle_{\ky=\pi} \langle \STRI \rangle_{\prod \ko \in (0,\pi)} \neq 0$. This OP would vanish in the other phases because the behavior of the $\ky=0,\pi$ DOF (i.e. viewed as topological or trivial Majorana chains) would be different. Regions where the order parameters take nonzero values are shown in Fig. \ref{fig:NLOP_plot}. 

\indent The fermion parity operator used in the bulk of the OPs given in the previous section can be reduced to parity for just the $\ky=0,\pi$ DOF, since the remaining momenta DOF always remain trivial in the $p+ip$ model. This reduction may not be applicable in general, as when interactions are added.  

\indent We emphasize that this model, as in the case of the single Kitaev Majorana chain, has additional symmetries beyond $\Ztwo \times \ZN$ which puts constraints on the construction of the OP. For instance, a choice of  $O_{L/R} \propto (\chi, \bchi)$ is different from $(\bchi, \chi)$, as we saw in the single chain. However, as discussed previously, for models which only have $\Ztwo \times \ZN$ symmetry, only these symmetries need to be accounted for; therefore, order parameters constructed using the general principles described work generically. For instance, $S_{3}, S'_{3}$ and similar order parameters are all suitable choices to distinguish Class 3 from the other symmetric fermionic phases. 

\subsection{Ising condensate as a model for nontrivial symmetric bosonic phase}
\indent To give an alternative picture of the quantum phases and the physical origin of the selection rules, we consider the nontrivial symmetric bosonic phase (Class 2) for the case of two chains (N=2). This is a ``Haldane-like" phase in that it is protected by a $\Ztwo \times \Ztwo$ symmetry. Our construction will be analogous to higher dimensional ones for symmetry-protected topological phases, in that two ``dual" defects, (i.e. one defect is nonlocal in the variables in which its partner is local) will be bound and then condensed. The composite object will carry a nontrivial quantum number under the symmetries. 

\indent Let $\sigma, \tau$ be two Ising variables with the $\Ztwo$ symmetries $\prod_{i} \sx(i), \prod_{i} \taux(i)$. Condensing domain walls of $\sigma$, created at site $j$ by $\prod_{i<j} \sx(i)$, would lead to an Ising disordered phase $\lang \sz(i) \rang = 0$, while the ordered phase is realized by condensing spin flips created by $\sz$. Consider condensing a bound state of a $\sigma$ defect and its $\tau$ dual defect, e.g. the composite object $\rho(j) = \prod_{i < j} \sx(i) \tauz(j)$. If no symmetries are broken, this will yield a topological phase. To preserve the symmetries, for instance, one can also condense $\tau$ domain wall and $\sigma$ spin flip pairs, $\delta(j) = \prod_{i \leq j} \taux(i) \sz(j)$. A string order parameter for this topological phase will be:
\begin{equation}
\begin{split}
\lang \rho(i) \rho(j) \delta(i) \delta(j) \rang &= \\
& \lang \tauz(i) \sy(i) \prod_{i < k < j} \sx(k)\taux(k) \tauy(j) \sz(j) \rang
\end{split}
\end{equation}

\indent The order parameter is of the general form discussed previously. It consists of applying one symmetry over the bulk (here,$\prod_{i} \sx(i)\taux(i)$) and terminating with operators ($\tauz \sy, \tauy \sz$) which are even under this symmetry and odd under the other symmetries $\prod_{i} \sx(i), \prod_{i} \taux(i)$.

\indent Based on this description, we write a Hamiltonian which realizes this topological phase. Consider starting at the critical point of a pair of decoupled Ising models: 
\begin{align}
H_{0} &= -\sum_{i} \left[ \sx(i) + \taux(i) + \sz(i) \sz(i+1) + \tauz(i) \tauz(i+1) \right]
\end{align}

\noindent and adding correlations for the $Z_{2}$ charge and domain wall bound objects in order to induce condensation of these composites: 
\begin{subequations}
\begin{align}
H_{1} &= -\sum_{i} \left[ \rho(i)\rho(i+1) + \delta(i)\delta(i+1) \right] \\
&= -\sum_{i} \left[ \sz(i)\sz(i+1)\taux(i+1) + \tauz(i)\tauz(i+1)\sx(i) \right]
\end{align}
\end{subequations}

\indent The Hamiltonian $H(\lambda) = H_{0} + \lambda H_{1}$ realizes the nontrivial topological phase for $\lambda > 1$ with nonlocal order parameter $\rho(i) \delta(i)$; this can be seen by making a dual transformation on one of the $\Ztwo$ variables and mapping onto the quantum Ashkin-Teller model.\cite{Kohmoto1, Kohmoto2} Moreover, $H_{1}$ itself is exactly solvable and its ground state is a so-called cluster state.\cite{Vedral} There is a four-fold degeneracy on a chain with sites 1 to L. On each edge, we can construct a spin-1/2 algebra with local operators; for instance, $\sz(1)\taux(1)$, $\sx(1)\tauz(2)$, and $\sy(1)\taux(1)\tauz(2)$ operate on the left edge while $\sx(L)\tauz(L)$, $\sz(L-1)\taux(L)$, and $\sz(L-1)\sx(L)\tauy(L)$ operate on the right edge. Since we can map within the ground state manifold via edge and not bulk operators, the distinction between the degenerate states is topological rather than associated with symmetry breaking.

\section{Conclusion}

\indent We have given string or brane-type nonlocal order parameters (Sec. \ref{sec:SecI}) to distinguish the phases of quasi-1D topological superconductors of spinless fermions with interactions and no symmetries. These order parameters measure fermion parity in their bulk and are terminated by fermionic or bosonic operators at their edges; we illustrated how they probe the different natures of the Majorana pairings in the topological and trivial phases. They would be an alternative way to detect Majorana fermions via quantum gas microscope measurements in cold atom systems (Sec. \ref{sec:SecII}). We also gave an example of an order parameter for two chains which only involves fermion parity and hence would be measurable using current experimental techniques but which detects a topological phase.

\indent The addition of translation to the system as a protecting symmetry (Sec. \ref{sec:SecIII}) distinguished among certain interesting 2D phases in the quasi-1D limit. We elaborated on how two 1D $\Ztwo$ invariants are distinguished by the classification in the case of free fermions; in particular, this allows us to distinguish the 2D Chern number mod 2, for instance the $p+ip$ strong topological superconductor and the weak topological superconductor.  

\indent We constructed simple general rules (Sec. \ref{sec:SecIV}) which the terminating operators of a nonlocal order parameter should satisfy in order to uniquely distinguish among the fermionic symmetric phases (four for $N$ even and two for $N$ odd), even in the case of interactions. We sketched a direct fermionic derivation of these rules (App. \ref{sec:AppIII}), which may be extendable to other symmetry classes. We illustrated the construction by giving a nonlocal order parameter for the $p+ip$ topological superconductor which distinguishes it from the weak topological superconductor or the trivial phase in the quasi-1D limit and which is robust to interactions. Attempts at extending string to brane order for coupled chains have been discussed in other contexts; here, we note that taking products of single chain string order parameters can work because of the additional protecting $\ZN$ symmetry. 

\indent In summary, we have devised uniquely identifying nonlocal order parameters for the symmetric phases of coupled topological superconducting spinless fermion chains with interactions. We considered both the case of (i) no protecting symmetries and (ii) transverse translation symmetry. The extension of our ideas to incorporate time reversal symmetry in 1D or to construct higher dimensional order parameters for fermionic topological phases would be interesting.\\

\section*{Acknowledgements}
\indent Y.B. thanks X. Chen, A. Turner, L. Fidkowski, and E. Altman for helpful conversations and L. Fidkowski for feedback on the manuscript. A.V. thanks E. Altman and E. Demler. Funding from NSF GRFP under Grant No. DGE 1106400 (Y.B.) and the Army Research Office with funding from the DARPA Optical Lattice Emulator program and NSF-DMR 0645691 (A.V.) is acknowledged.

\appendix
\renewcommand{\thesection}{\Alph{section}}
\section{Applying a broken symmetry on a domain}
\label{sec:AppI}
\indent Consider a quasi-1D spin system with an infinite dimension indexed by $j$ and which has a discrete broken symmetry operator $u=\prod_{j=-\infty}^{\infty} u_{j}$. We argue that any state $|\psi_{0}\rangle$ in the ground state manifold obeys \begin{math} \lim_{N \rightarrow \infty} \langle \psi_{0} |\prod_{j=-N}^{N} u_{j} |\psi_{0} \rangle \rightarrow 0 \end{math}, with a special ordering of the limits. It applies even when mapped to fermions because it considers arbitrary ground state choices.

\indent Let $\{ |\eta_{i}\rangle \}_{i=1}^{M}$ be the broken symmetry states which are mapped to each other under $u$. We explain that $\lim_{N \rightarrow \infty} \langle \eta_{i} | \prod_{j=-N}^{N} u_{j} |\eta_{k} \rangle \rightarrow 0$ for any $i,k$. If $i=k$, $u$ creates a finite-sized domain which is orthogonal to the original state as the domain size increases $N \rightarrow \infty$. For instance, for the quantum Ising model with a $\Ztwo$ broken symmetry, $\lim_{N \rightarrow \infty} \langle \uparrow, \downarrow | \prod_{j=-N}^{N} \sigma^{x}_{j} |\uparrow, \downarrow \rangle \rightarrow 0$, where $|\uparrow\rangle, |\downarrow\rangle$ denote the broken symmetry states in the thermodynamic limit. Off-diagonal matrix elements $i \neq k$ also vanish due to the order of our limiting procedures; since the thermodynamic limit precedes $N\rightarrow \infty$ there is always an infinite region outside the domain $\left[-N,N \right]$ where the broken symmetry states are orthogonal. Practically, this means that the system size must be much larger than the domain over which the broken symmetry is applied in order to yield an asymptotically vanishing value. We expect that our description can be formalized with matrix product states $\{ A^{i}_{\alpha} \}_{\alpha=1}^{d}$ ($A^{i}_{\alpha} = \Gamma^{i}_{\alpha} \Lambda^{i}$ in the notation used in the main text) associated with $|\eta_{i}\rangle$ and on-site physical dimension $d$ by considering the eigenproblem of the transfer matrices $\mathbb{E}^{ij} = \sum_{\alpha=1}^{d} A^{i}_{\alpha} \otimes (A^{j}_{\alpha})^{*}$, $1 \leq i,j \leq M$, which govern the behavior of state overlaps.

\section{Fermionic classification}
\label{sec:AppII}

\indent We follow the approach developed in [\!\!\!\citenum{TurnerClass}] for 1D fermionic and bosonic systems. We consider the system $\Omega$ with periodic boundary conditions and a unique gapped ground state, and partition $\Omega = \Omega_{S} \cup \Omega_{E}$ into a subsystem $\Omega_{S}$ and the environment $\Omega_{E}$. Let an observable be $O$. Consider the effective action $\hat{O}$ of this operator in the space spanned by the low entanglement energy (EE) Schmidt states obtained from the ground state on subsystem $\Omega_{S}$. Ref.\citenum{TurnerClass} observed that the action reduces to that of two operators $O_{L}, O_{R}$ acting locally near the left and right edges, respectively, of $\Omega_{S}$, i.e. $\hat{O} \sim O_{L} O_{R}$. That is, in this subspace spanned by low EE states, states are distinguished by physics near their edges (as observables have ``fractionalized" into two spatially separated pieces) but behave similarly in their bulk. Symmetry-protected phases are distinguished by the commutation relations obeyed by the edge operators. 

\indent Our two $\Ztwo \times \ZN$ commuting symmetry generators are parity and translation P, T. They fractionalize as $\hat{P} \sim P_{L} P_{R}$ and $\hat{T} \sim T_{L} T_{R}$. We fix $\hat{P}^{2} = P^{2}_{L} = P^{2}_{R} = 1$ and $\hat{T}^{N} = T^{N}_{L} = T^{N}_{R} = 1$. Define angles $\mu, \mu'$ with $P_{L} P_{R} = e^{i\mu} P_{R} P_{L}$, $T_{L} T_{R} = e^{i\mu'} T_{R} T_{L}$ which are $0, \pi$ since fractional pieces can be fermionic or bosonic. 

\indent We claim that $\mu, \mu'$, along with an additional assumption that $\hat{P} \TL = e^{i\mu'} \TL \hat{P}$, are sufficient to distinguish the quantum phases, since the other commutations follow from these. The complete operator $P$ determines whether operators such as $\TL$ are bosonic or fermionic (value of $\mu'$) and it is natural to assume that its effective form does also. Parity is in this way a more fundamental operator for fermionic systems compared to other symmetries. Other commutation relations follow, such as $\PL \hat{T} = \hat{T} \PL e^{i\mu'}$.

\indent An equation such as $\PL \hat{T} = \hat{T} \PL e^{i\mu'}$ imposes a constraint since $\PL^{2} = \hat{T}^{N} = 1$; namely, $\mu' = \pi$ is not allowed if $N$ is odd. Hence, we recover the same symmetric fermionic phases as we would by mapping the bosonic group cohomology classification to fermions: $(\Ztwo)^{2}$ for $N$ even and $\Ztwo$ for $N$ odd. We additionally have a direct fermionic description of the phases based on effective forms of symmetry operators. Finally, to establish a correspondence between the bosonic and fermionic descriptions, we should understand when the Jordan-Wigner mapped versions of the fermionic symmetry operators are broken or unbroken in the bosonic variables. This leads us to find that parity is broken when $\mu = \pi$, while translation is broken when $\mu'=\pi$ \emph{and} $\mu=\pi$; this is summarized in Table \ref{tab:Correspondence}.

\section{Sketch of proof for fermionic selection rules}
\label{sec:AppIII}

\indent We sketch a proof that the terminating operators should satisfy certain transformation (selection) rules in order for the nonlocal order parameter to remain nonzero in one symmetric fermionic phase and vanish in the others; the result is Eqs. \ref{eq:SelectionRule}. We will evaluate the long-distance limit of the string or brane OP $\lang \oL \left[ \prod_{j \in \OmS} \Sigma_{j} \right] \oR \rang$ in the ground state, with $\oL, \oR$ local terminating operators and $\Sigma_{j}$ an on-site symmetry. The asymptotic form of a nonlocal order parameter in a symmetric phase is really a two-point function of certain operators because symmetries reduce to acting on the edges of the domain over which they are applied. We use the effective forms for fermionic symmetries from the fermionic classification; though they are state dependent, we only rely on properties of the phases. \\
\indent We consider as in App. \ref{sec:AppII} a closed system $\Omega$ partitioned into a subsystem $\Omega_{S}$ over which the symmetry $\Sigma$ acts and an environment $\Omega_{E}$, on whose edges $\oL, \oR$ act. The ground state has Schmidt decomposition $|\psi\rang = \sum_{a} e^{-E_{a}} |\phi_{a}\rang |\eta_{a}\rang$ where $\phi_{a}, \eta_{a}$ are for $\Omega_{S}, \Omega_{E}$, respectively. We specialize to the case of interest where fermion parity $P_{\OmS}$ is applied in the bulk. The idea of [\!\!\citenum{TurnerClass}] is that $\lang \phi_{a} | \PS |\phi_{a'} \rang \approx \lang \phi_{a} | \PSL \PSR |\phi_{a'} \rang$ (with effective forms $\PSL, \PSR$ on $\Omega_{S}$) for states with low entanglement energy (EE), so that $a,a' < \chi$ with $\chi$ a cutoff. The forms $\PSL, \PSR$ are localized to a distance $l$ near the edges of $\Omega_{S}$ which increases with $\chi$. While the replacement by effective forms is approximate, it is good because states with high EE contribute less to evaluations of observables. Hence:
\begin{align*}
&\lang \psi | \oL \PS \oR |\psi \rang \approx \\ 
\sum_{a,a'<\chi} &e^{-E_{a}-E_{a'}} \lang \phi_{a} | \PSL \PSR |\phi_{a'} \rang \lang \eta_{a} | \oL \oR |\eta_{a'} \rang\\ 
 = \sum_{a,a'<\chi} &e^{-E_{a}-E_{a'}} \lang \phi_{a}| \lang \eta_{a}|\oL \PSL \PSR \oR |\phi_{a'}\rang |\eta_{a'}\rang \\
&\equiv \lang \tilde{\psi} | \oL \PSL \PSR \oR |\tilde{\psi}\rang \tag{26}
\end{align*}
\addtocounter{equation}{1}

\noindent where $|\tilde{\psi}\rang \equiv \sum_{a < \chi} e^{-E_{a}} |\phi_{a}\rang|\eta_{a}\rang$ is a good approximation to ground state $\psi$. We first take the thermodynamic limit of the closed system and then $\OmS$ so that the evaluations at the left and right boundaries of $\OmS$ near $\OmE$ decouple; the nonlocal order parameter reduces to an evaluation of local operators:
\begin{align}
&\lang \psi | \oL \PS \oR |\psi \rang \approx \lang \tilde{\psi} | \oL \PSL |\tilde{\psi}\rang \lang \tilde{\psi} | \PSR \oR |\tilde{\psi}\rang
\end{align}

\noindent We then take the limit $\chi, l \rightarrow \infty$, so $\PSL,\PSR$ penetrate further into the bulk of the (infinite) subsystem S, $|\tilde{\psi}\rang \rightarrow |\psi \rang$, and the approximation improves. 

\indent Consider the transformation properties of just one edge evaluation, for instance $\lang \tilde{\psi} | \oL \PSL |\tilde{\psi} \rang$. $\PSL$ has known transformation rules under the symmetries which are characteristic of the quantum phase. How must $\oL$ transform in order to force the expression to vanish? $|\tilde{\psi}\rang$ is approximately an eigenstate of the effective forms of the total symmetries $P_{\Omega},T_{\Omega}$, becoming exact in the above limits. Consider introducing translation, for instance:
\begin{align}
\lang \psi | \oL \PSL |\psi\rang &= \lang \psi | T_{\Omega}^{\dag} \oL \PSL T_{\Omega} |\psi\rang \\
T_{\Omega}^{\dag} \oL \PSL T_{\Omega} &= (\TE^{\dag} \oL \TE) (\TS^{\dag} \PSL \TS)
\end{align}

\indent (Note that $\TS, \PS$ are bosonic). From App. \ref{sec:AppII}, we have $\TS^{\dag} \PSL \TS = e^{i\mu'} \PSL$. In order to have $\lang \oL \PSL \rang \neq 0$, we need $\TE^{\dag} \oL \TE = e^{-i\mu'} \oL$. Applying the same argument with parity symmetry and using $\PS^{\dag} \PSL \PS = e^{i\mu} \PSL$ implies $\PE^{\dag} \oL \PE = e^{-i\mu} \oL$ is needed also. When the terminating operator $\oL$ satisfies transformation laws different from the one characterizing the quantum phase $(\mu,\mu')$ of the system, the local evaluation $\lang \psi | \oL \PSL |\psi \rang$ will vanish asymptotically. In summary, we need:
\begin{subequations}
\label{eq:SelectionRule}
\begin{align}
P^{\dag} \oL P &= e^{-i\mu} \oL \\
T^{\dag} \oL T &= e^{-i\mu'} \oL
\end{align} 
\end{subequations}

\noindent so that the nonlocal order parameter vanishes in the fermionic symmetric phases characterized by angles different from $(\mu,\mu')$. These selection rules support the conclusions reached using bosonic selection rules and local order parameters for bosonic symmetry breaking. For instance, a nonlocal order parameter for Class 2 $(\mu,\mu')=(0,\pi)$ should have $\oL,\oR$ chosen to be even under fermion parity and odd under translation, as described also in the main text.

\bibliographystyle{apsrev}  
\bibliography{NLOP_refs_Sept18} 

\end{document}